\documentclass{pasj00}
\draft

\begin{document}
\SetRunningHead{Dogiel et al.}{Neutral Iron Line from Sgr~B2 by Subrelativistic Protons}
\Received{2000/12/31}
\Accepted{2001/01/01}
\title{
K-shell Emission of Neutral Iron Line from Sgr B2 Excited by
Subrelativistic Protons.}

\author{Vladimir \textsc{Dogiel}$^{1}$, Dmitrii {\sc Chernyshov}$^{1,3}$,
Katsuji {\sc Koyama}$^{2}$, Masayoshi {\sc Nobukawa}$^{2}$  and
Kwong-Sang {\sc Cheng}$^{3}$}
 \affil{$^1$P.N.Lebedev Institute, Leninskii
pr, 53, 119991 Moscow, Russia}
\email{dogiel@lpi.ru}
 \affil{$^2$Department of Physics, Graduate
school of Science, Kyoto University, Sakyo-ku, Kyoto 606-8502}
\affil{$^3$Department of Physics, University of Hong Kong,
Pokfulam Road, Hong Kong, China}


%


\maketitle

\begin{abstract}
We investigated the emission of K$\alpha$ iron line from  the
massive molecular clouds in the Galactic center (GC). We assume
that at present the total flux of this emission consists of time
variable component generated by primary X-ray photons ejected by
Sagittarius A$^\ast$ (Sgr A$^\ast$) in the past and a relatively
weak quasi-stationary component excited by impact of protons which
were generated by star accretion onto the central black hole. The
level of  background  emission was estimated from
  a rise of the 6.4 keV line intensity in the direction of several molecular clouds,
  that
we interpreted as a stage when the X-ray front ejected by Sgr
A$^\ast$ entered into these clouds. The 6.4 keV emission before
this intensity jump we interpreted  as  emission generated by
subrelativistic cosmic rays there. The cross-section of K$\alpha$
vacancies produced by protons differs from that of electrons or
X-rays. Therefore, we expect that this processes can be
distinguished from the analysis of the equivalent width of the
iron line and  time variations of the width can be predicted. The
line intensity from the clouds depends on their distance from Sgr
A$^\ast$ and the coefficient of spacial diffusion near the
Galactic center. We expect that in a few years the line intensity
for the cloud G\,0.11$-$0.11 which is relatively close to Sgr
A$^\ast$ will decreases to  the level $\lesssim$ 10\% from its
present value. For the cloud Sagittarius B2 (Sgr B2) the situation
is more intricate. If the diffusion coefficient $D\gtrsim 10^{27}$
cm$^2$~s$^{-1}$ then the expected stationary flux should be about
10\% of its level in 2000. In the opposite case the line intensity
from Sgr B2 should drop down to zero because the protons do not
reach the cloud.

\end{abstract}

\section{Introduction}
The bright iron fluorescent K$\alpha$ line in the direction of the
molecular clouds  in the Galactic center (GC) region was predicted
\citep{suny} and then discovered \citep{koya1} more than twenty
years ago. It was assumed that this flux arose due to the
K-absorption of keV photons by dense molecular clouds irradiated
by external X-rays, possibly from the super-massive black hole,
Sagittarius A$^\ast$ (Sgr A$^\ast$), which was active in the
recent past, ($300$ -- $400$ years ago  \citep{suny,koya1}), but
is almost unseen at present (see  e.g.  \cite{bag} and
\cite{porquet}). Recent observations found a steady decrease of
the X-ray flux from Sagittarius~B2 (Sgr~B2)
\citep{koya08,inui,terrier,nobukawa1}. This is a strong evidence
that the origin of the variable component is, indeed, a reflection
of the primary X-ray flare.

 The duration of Sgr A$^\ast$ activity is uncertain. Thus,
\citet{mura}  obtained the luminosity history of the galactic
nuclei Sgr A$^\ast$ during the last 500 years. They concluded that
Sgr A$^\ast$ was as luminous as $F_{fl}\sim 10^{39}$ erg s$^{-1}$
 a few hundred years ago, and has dimmed gradually since
then. \citet{rev1}  found no significant variability of the line
flux from Sgr B2 during the period 1993–-2001. The constancy of
the line flux meant that the luminosity of Sgr A$^\ast$ remained
approximately constant for more than 10 years a few hundred years
ago. \citet{inui} confirmed this  variability of Sgr A$^\ast$
activity with a time scale $\sim$10 years. And, finally
\citet{ponti} concluded  that this activity might have started a
few hundreds of years ago and lasted until about 70 -- 150 years
ago.

An appropriate duration of Sgr A$^\ast$ X-ray activity can be
caused by shocks resulting from interaction of  jets with the
dense interstellar medium \citep{yu}.

K$\alpha$ emission from the clouds can be generated by
subrelativistic electrons with energies above 7 keV.  This model
was proposed by \citet{yus1} (see also \cite{yus2,yuz}) who
assumed that a correlation between the nonthermal radio filaments
and the X-ray features when combined with the distribution of
molecular gas might suggested that the impact of the
subrelativistic electrons with energies 10--100 keV from local
sources with diffuse neutral gas produced both bremsstrahlung
X-ray continuum  and diffuse 6.4 keV line emission. The excess of
supernova remnants detected in the GC region was supposed to be
responsible for enhancing the flux of subrelativistic electrons.
The characteristic time of K$\alpha$ emission in this case is
about $\geq 1000$ years, i.e. about the lifetime of
subrelativistic electrons  (for the rate of electron energy losses
see e.g. \cite{haya}). The total energy release of a supernova is
about $10^{51}$ erg.

Observations indicated on even more energetic phenomena which
might occur in the GC. Thus, a hot plasma with the temperature
about 10 keV was found in the GC which can be heated if there are
sources with a power $\sim 10^{41}$ erg s$^{-1}$ (see e.g.
\cite{koya1}), which could be generated by  events of huge energy
release in the past. It was shown that the energy about $10^{53}$
erg can be released if the central black hole captured a star (see
e.g. \cite{alex05,cheng1,cheng2}). As a result, a flux of
subrelativistic protons is ejected from the GC, which heats the
central region \citep{dog_pasj}. These protons can also produce
6.4 keV line emission from molecular clouds \citep {dog_pasj1},
which is, however, stationary because the lifetime of these
protons $\tau_p\sim 10^7$ yr \citep{dog_aa} is much longer than
the characteristic time of star capture by the central black hole
($\tau_c\sim 10^5$ yr) \citep{alex05}.  This scenario assumed at
least two components of the X-ray line and continuum emission from
the clouds: the first is a time variable component generated by
X-rays  from  sources in the GC, and the second is a
quasi-stationary component produced  by  subrelativistic protons
interacting with the gas.

 The question whether  the X-ray emission from the central region
(within $\leq \timeform{0.3D}$ radius) is really diffuse was
analysed in \citet{koya2} who showed that the hot plasma
distribution in the GC, traced by the 6.7 keV iron line emission,
did not correlate with that of point sources whose distribution
was derived from IR observations that differed from the other disk
where the correlation was quite good. Recently, \citet{rev09}
showed from the Chandra data that most ($\sim 88$\%) of the ridge
emission is clearly explained by dim and numerous point sources.
Therefore, at least in the ridge emission, accreting white dwarfs
and active coronal binaries are considered to be main emitters. We
notice however that \cite{rev09} observed regions in the disk
located at $\timeform{1.5D}$ away from the GC.

Observations of the 6.4~keV flux from Sgr B2 have not found up to
now any reliable evident stationary component  though as predicted
by \citet{ponti} a fast decrease of 6.4 keV emission observed with
XMM-Newton for several molecular clouds  suggested that the
emission generated by low energy cosmic rays, if present, might
become dominant in several years.  Nevertheless, for several
clouds, including Sgr B, observations show temporary variations of
6.4 keV emission both rise and decay of intensity (see
\cite{inui,ponti}). We interpreted this rise of emission as a
stage when the X-ray front ejected by Sgr A$^\ast$ entered into
these clouds and the level of background generated by cosmic rays
as the 6.4 keV emission before the intensity jump.

Below we shall show that if this stationary component exists it
can be predicted from time variations of the line emission from
the clouds.

\section{Equivalent Width of the 6.4 keV Line}
In the framework of the reflection model, primary X-rays from an
external source produce fluxes of continuum and line emission from
irradiated molecular clouds.  In principle, the surface brightness
distribution, the equivalent width and the shape of the
fluorescent line depend on the geometry of the
source-reflector-observer (see \cite{suny1}) but for rough
estimates we can neglect this effect. The continuum flux from the
clouds is proportional roughly to
\begin{equation}
F_X\propto n_H\sigma_Tc N_X\,,
\end{equation}
where $n_H$ if the hydrogen density in the cloud, $\sigma_T$ is
the Thomson cross-section, and $N_X$ is the total number of
primary photons with the energy of a produced X-ray $E_X\sim 7.1$
keV inside the cloud.

The flux of 6.4 keV line is
\begin{equation}
F_{6.4}\propto n_H\sigma^X_{6.4}c \eta N_X\,,
\end{equation}
where $\eta$ is the iron abundance in the cloud and
$\sigma^X_{6.4}$ is the cross-section of the line production by
the primary X-ray flux. Then the equivalent width ({\it eW}) of
the line in the framework of the reflection model is
\begin{equation}
eW=\frac{F^X_{6.4}}{F_X(E_X=6.4~keV)}\propto
\frac{\sigma^X_{6.4}\eta}{\sigma_T}=f(\eta)\,.
\end{equation}

The intensity of the Fe K$\alpha$ line excited by subrelativistic
particles (electrons or protons) in a cloud can be calculated from
\begin{equation}
F_{K_\alpha}=4\pi \eta\omega_K \int\limits_rn_H(r)
r^2dr\int\limits_E v(E)\sigma_K \tilde{N}(E,r)dE\,,
\end{equation}
where $v$ and $E$ are the velocity and the kinetic energy of
subrelativistic particles, $\sigma_{K}$ is the cross-section for
6.4 keV line production by subrelativistic particles,
\begin{equation}
\sigma_K=\sigma_Z^I\eta\omega_Z^{KI}\,.
\end{equation}
Here $\sigma_Z^I$ is the cross section for the K-shell ionization
of atom Z by a charged particle of energy $E$ (see \cite{garcia,
quarl}), $\omega_Z^{KI}$ is the K{\it i} fluorescence yield for an
atom Z.

The flux of  bremsstrahlung radiation is
\begin{equation}
\Phi_x=4\pi\int\limits_{0}^\infty ~n_H(r) r^2dr\int\limits_E
~dE\tilde{N}(E,x,t){{d\sigma_{br}\over{dE_x}}}v(E)\,. \label{i_br}
\end{equation}
Here $d\sigma_{br}/{dE_x}$ is the cross section of bremsstrahlung
radiation (see \cite{haya})
\begin{equation}
  {{d\sigma_{br}\over{dE_x}}}={8\over 3}{Z^2}{{e^2}\over{\hbar c}}\left({{e^2}
  \over{m{c^2}}}\right)^2{{m{c^2}}\over{E^\prime}}{1\over{E_x}}
\ln{{\left(\sqrt{E^\prime}+\sqrt{{E^\prime}-{E_x}}\right)^2}\over{E_x}}\,,
\label{sbr}
\end{equation}
where  $E^\prime=E_e$ for electrons and
$E^\prime=E_p=(m_p/m_e)E_e$ for protons. One can find also all
these cross-sections in \citet{tatis}.

In principle the particle and X-ray scenarios can be distinguished
from each other from  characteristics of  emission from the clouds
because the  cross-sections for collisional and photoionization
mechanisms are quite different. If the photoionization
cross-sections are steep functions of energy (they vary
approximately as $E_X^{-3}$  from ionization thresholds), the
cross sections for collisional ionization have a much harder
energy dependence. Therefore, while fluorescence is essentially
produced by photons with energy contained in a narrow range of a
few keV above the ionization threshold, subrelativistic particles
can produce significant X-ray line emission in an extended energy
range above the threshold \citep{tatis}. The continuum emission in
these two models is also generated different processes : by the
bremsstrahlung in the collisional scenario and by the Thomson
scattering in the photoionization scenario.

\begin{figure}[h]
\begin{center}
\FigureFile(80mm,80mm){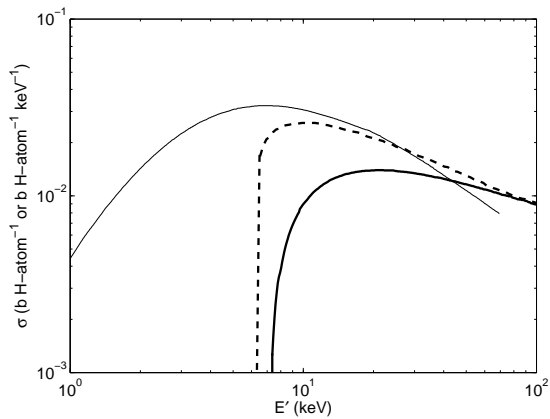}
\end{center}
\caption{ Cross section of electron and proton bremsstrahlung
radiation at the energy 6.4 keV, $d\sigma_{br}/{dE_x}$ (dashed
line), and the cross-sections $\sigma_K$ of K$\alpha$ production
for electron (thick solid line) and proton (thin solid line). Here
$E^\prime=E_e$ for electrons and $E^\prime=(m_e/m_p)E_p$ for
protons. Here $\omega_K$ equals
 0.3 and $\eta$  is taken  to be twice solar.The data for this figure was kindly sent to us by
Vincent Tatischeff.}\label{cr_sec}
\end{figure}

The cross-sections of bremsstrahlung and K$\alpha$ production by
subrelativistic protons and electrons are shown in figure
\ref{cr_sec}.
As one can see from the figure the cross-section of the proton
bremsstrahlung with the energy $E_p=(m_p/m_e)E_e$  is completely
the same as for electrons with the energy $E_e$ and for protons ,
as shown in figure~\ref{cr_sec} by the dashed line. However the
cross-sections $\sigma_K$ of K$\alpha$ lines produced by electrons
(thick solid line) and by protons (thin solid line) are quite
different.  If for electrons the cross-section $\sigma_K$ of the
iron line has a sharp cut-off at $E=7.1$ keV, that for protons is
rather smooth, and a contribution from protons with relatively
small energies can be significant.

The photoionization and collisional scenarios
 can be distinguished from
 the equivalent width of iron line. The {\it eW} depends on the chemical abundance in the GC, which is poorly
known for the GC. Direct estimations of the iron abundance there
provided by the  Suzaku group \citep{koya2,koya09} gave the value
from 1 to 3.5 solar. \citet{rev1} got the iron abundance for the
cloud Sgr B2 at about 1.9 solar. \citet{nobukawa} found that the
equivalent width  of line  emission from a cloud near Sgr A
requires the abundance higher than solar. For the line emission
due to impact of subrelativistic electrons, the iron abundance in
Sgr B2 should be about $4$  solar, while the X-ray scenario
requires $\sim 1.6$ solar. Therefore, \citet{nobukawa} concluded
that the irradiating model seemed to be more attractive than the
electron impact scenario. This abundance is compatible with the
value $\eta=1.3$ solar estimated by \citet{nobukawa1} from the iron
absorption edge at 7.1 keV.

The  {\it eW} for the case of particle impact depends on their
spectrum. Its value for power-law spectra of particles ($N\propto
E^{\gamma}$) is a function of the spectral index $\gamma$ and the
abundance $\eta$:
\begin{equation}
{\it eW}=\eta\omega_K\frac{ \int\limits_E v(E)\sigma_K(E)
E^{\gamma}dE}{\int\limits_ E ~E^{\gamma}({d\sigma_{br}({\bar
E},E)/{dE_x}})v(E)~dE}=f(\eta, \gamma)\,.
\end{equation}
For the solar iron abundance   the {\it eW} for electrons and
protons is shown in figure \ref{eqw_gamma}. It was assumed here
that the proton spectrum has a cut-off ($N=0$ at $E>E_{inj}$, see
below).
\begin{figure}[h]
\begin{center}
\FigureFile(80mm,80mm){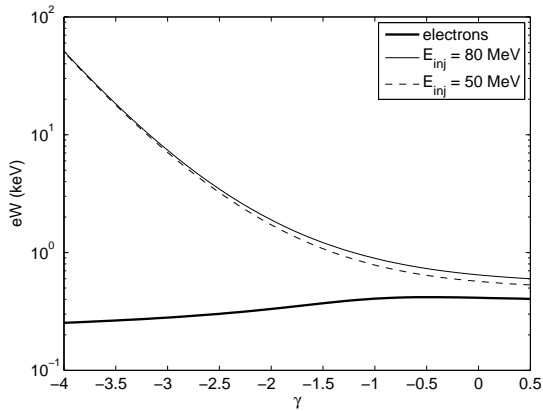}
\end{center}
\caption{Equivalent width of K$\alpha$ line  for the solar
abundance
  produced by electrons (thick solid line) and protons
(thin solid line for injection energy $E_{inj}=80$ MeV, dashed
line for injection energy $E_{inj}=50$ MeV) as a function of the
spectral index $\gamma$. }\label{eqw_gamma}
\end{figure}

One can see that the equivalent width of K$\alpha$ line generated
by electrons depends weakly on $\gamma$, and it varies from $\sim
250$ eV for soft spectra to $\sim 500$ eV for hard electron
spectra (see also in this respect \cite{yus2}). In the case of
protons the width variations are significant reaching its maximum
for very soft proton spectra. As one can see from this figure the
equivalent width weakly depends on the maximum energy of protons,
$E_{inj}$.

 Sources of high energy particles in the Galaxy generate quite
wide range of  characteristics of their spectra, though the most
effective process in the cosmic space, acceleration by shocks,
provide particle spectra with the spectral index $\gamma$ close to
-2. For the case of accretion we approximated the spectrum of
proton injection by the delta-function distribution which was
modified then by Coulomb losses into a power-law spectrum with
$\gamma=0.5$ (see \cite{dog_aa}). We notice, however, that this
delta-function approximation is a simplification of the injection
process.  As it was shown by \citet{chech} for jets, at first
stages of evolution the jet material moves by inertia. Then
because of excitation of plasma instabilities in the flux, the
particle distribution functions, which were initially delta
functions both in angle and in energy, transform into complex
angular and energy dependencies.

Below we present briefly  parameters of the proton spectrum for
the case of a star capture by a massive black hole (for details
see \cite{dog_pasj1, dog_aa}).

\section{Model of Proton Injection in the GC}

We mention first of all that penetration of subrelativistic
protons into  molecular clouds is supposed to be a rather natural
process in the Galaxy. Thus, investigations showed that heating
and ionization of Galactic molecular clouds can be produced by
subrelativistic protons penetrating there (see e.g.
\cite{dal:72,spit:75,na:94,dog_pasj1,crocker1}). If so, one expect
also a flux of the 6.4 keV line and continuum emission from these
clouds generated by these protons.

In the GC subrelativistic protons can be generated  by processes
of star accretion on the super-massive black hole. Once passing the
pericenter, the star is tidally disrupted into a very long and
dilute gas stream. The outcome of tidal disruption is that some
energy is extracted out of the orbit to unbind the star and
accelerate the debris. Initially about 50\% of the stellar mass
becomes tightly bound to the black hole, while the remainder 50\%
of the stellar mass is forcefully ejected \citep{ayal}. Then the
total number of subrelativistic protons produced in each capture
of one solar mass star is $N_k\simeq 10^{57}$.

The kinetic energy carried by the ejected debris is a function of
the penetration parameter $b^{-1}=r_t/r_p$, where $r_p$ is the
periapse distance (distance of closest approach) and $r_t$ is the
tidal radius. This energy  can significantly exceed that released
by a normal supernova ($\sim 10^{51}$~erg) if the orbit is highly
penetrating \citep{alex05},
\begin{equation}\label{energy}
  W\sim 4\times 10^{52}\left(\frac{M_\ast}{M_\odot}\right)^2
  \left(\frac{R_\ast}{R_\odot}\right)^{-1}\left(\frac{M_{\rm bh}/M_\ast}{10^6}\right)^{1/3}
  \left(\frac{b}{0.1}\right)^{-2}~\mbox{erg}\,.
\end{equation}
where $M_\ast$ and $R_\ast$ is the mass and the radius of the
captured star, and $M_{\rm bh}$ is the mass of black hole.

For the star capture time $\tau_s\sim 10^{4}-10^5$ years
\citep{alex05}
 it gives a power input $W \sim  10^{41}$ erg s$^{-1}$.
 The mean
kinetic energy per escaping nucleon is given by
\begin{equation}\label{esc}
  E_{\rm inj}\sim 42 \left(\frac{\eta}{0.5}\right)^{-1} \left(\frac{M_\ast}{M_\odot}\right)
  \left(\frac{R_\ast}{R_\odot}\right)^{-1}\left(\frac{M_{\rm bh}/M_\ast}{10^6}\right)^{1/3}
  \left(\frac{b}{0.1}\right)^{-2}~\mbox{MeV}\,,
\end{equation}
where $\eta M_\ast$ is the mass of escaping material. For the
black-hole mass $M_{\rm bh}=4.31 \times 10^6~M_{\odot}$ the energy
of escaping particles is
\begin{equation}
E_{\rm inj} \sim 68 (\eta /0.5)^{-1} (b/0.1)^{-2}~\mbox{MeV
nucleon$^{-1}$}
\end{equation}
 when a one-solar mass star is
captured.

 Subrelativistic protons lose their energy by collision with background
 particles and the lifetime of subrelativistic protons in  the GC with
 energies $E\leq 100$ MeV is
of the order of
\begin{equation}
\tau_{p} \sim \sqrt{\frac{E_p^3}{2m_p}} \frac{m_e}{3\pi
ne^4\ln\Lambda}\sim  10^7~\mbox{years}
\end{equation}
where $n\sim 0.1$ cm$^{-3}$ is the plasma density in the GC, $e$
and $m$ are the electron charge and its rest mass, respectively,
and $ln\Lambda$ is the Coulomb logarithm. Because
$\tau_s\ll\tau_{p}$, then the  proton injection can be considered
as quasi-stationary.

The spatial and energy distribution of these protons in the
central GC region can be calculated from the equation
\begin{equation}\label{pr_state}
  \frac{\partial N}{\partial t}  - \nabla \left(D\nabla N \right)+
  \frac{\partial}{\partial E}\left( \frac{dE}{dt} N\right) =
  Q(E,t)\,,
\end{equation}
where $dE/dt$ is the rate of  Coulomb energy losses,  $D$ is the
spatial diffusion coefficient in the intercloud medium and the rhs
term $Q$ describes the process proton injection in the GC
\begin{equation}
  Q(E, {\bf r}, t) = \sum \limits_{k=0}N_k\delta(E-E_{inj})\delta(t - t_k)\delta({\bf r})\,,
\end{equation}
where $N_k$ is the number of injected protons and  $t_k=k \times
T$ is the injection time.

The proton distribution inside molecular clouds is described by
similar equation but with a different diffusion coefficient and
rates of energy losses
\begin{equation}\label{pr_cl}
 \frac{\partial}{\partial E}\left( b_c(E) \tilde{N}\right) -  D_c\frac{\partial^2}{\partial x^2} \tilde{N}
 = 0\,,
\end{equation}
with the boundary conditions
\begin{equation}
\tilde{N}|_{x=0}=N_c,~~~~~~~~~~~~\tilde{N}_p|_{x=\infty}=0\,.
\end{equation}
where $N_c$, the proton density at the cloud surface, is
calculated with equation~(\ref{pr_state}), $D_c$ and $b_c$ are the
diffusion coefficient and the rate of energy losses inside the
cloud.  The value of $D_c$ for the clouds is uncertain though
there are theoretical estimates of this value provided by
\citet{dog87} who gave the value $\sim 10^{24}-10^{25}$
cm$^2$~s$^{-1}$. For details of calculations see
\citet{dog_pasj1}.

\section{Stationary and Time-Variable Components of X-Ray Emission from the GC Molecular Clouds}
The following analysis is based on the cloud observations by
XMM-Newton obtained by \citet{ponti}. These clouds showed
different time variations of the line emission which were
interpreted by the authors in terms of the reflection model. The
distances
 of the clouds from Sgr A$^\ast$ was  chosen in that way to
 explain the observed variations of the line emission for each of these clouds. Several
 clouds of the Bridge complex show a rather low flux before a sudden jump of the 6.4 keV intensity
in about one order of magnitude that was interpreted as a result
of the X-front radiation which just had reached these clouds (see
figure~5, 6, and 11 in \cite{ponti}). Basing on these observations we
make the two key assumptions:
\begin{enumerate}
 \item The low level of 6.4 keV intensity from the clouds before the jump represents
 a stationary component of the emission from the clouds. This assumption does not seem to be
incredible. Suzaku observations show also faint 6.4 keV emission
from the GC region
 which is more or less uniformly distributed there (see \cite{koya09}). We cannot exclude that
 this extended diffuse emission may also represent a stationary line
 component.
\item This emission from the clouds before the jump is generated by proton impact.
\end{enumerate}

For our analysis we used also parameters of the two other clouds
which showed time variations of the 6.4 keV emission. With some
modeling of proton penetration into the clouds described in the
previous section we can calculate stationary components of
continuum and the 6.4 keV line emission from the  clouds produced
by protons. The diffusion coefficient $D$ in the GC is unknown and
therefore is a free parameter of the problem. For calculations we
took parameters for the three clouds which are listed in
\citet{ponti}:
\begin{itemize}
\item Bridge, the density $n_H=1.9\cdot 10^4$ cm$^{-3}$, the radius of the cloud $r = 1.6$ pc,
the distance from Sgr A$^\ast$ $R = 63$ pc;
\item the same for the cloud G\,0.11$-$0.11, $n_H=1.8\cdot 10^{3}$ cm$^{-3}$, $r = 3.7$ pc, $R = 30$
pc;
\item the same for Sgr B2, $n_H = 3.8\cdot 10^4$ cm$^{-3}$, $r = 7$ pc, $R =
164$ pc.
\end{itemize}

\begin{figure}[ht]
\begin{center}
\FigureFile(80mm,80mm){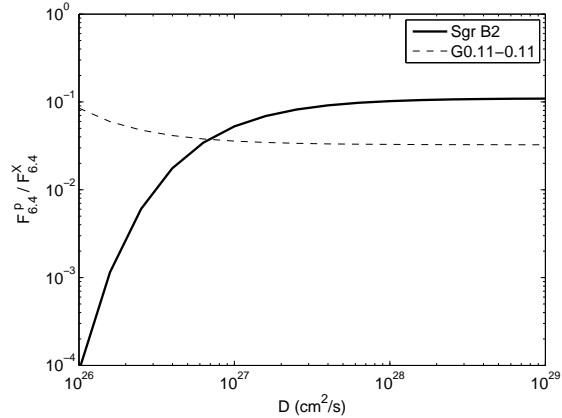}
\end{center}
\caption{The ratio $F^p_{6.4}/F^X_{6.4}$ for the cloud G\,0.11$-$0.11
and Sgr B2 as a function of the diffusion coefficient $D$}
\label{fratio}
\end{figure}

Intensity of the 6.4 keV line produced by photoionization depends
on the number of primary photons penetrating into a cloud. The
density of primary X-ray flux from Sgr A$^\ast$
 decreases with the distance $R$ as: $\propto
R^{-2}$. Then with the known parameters of X-ray and proton
production by Sgr A$^\ast$ we can calculate for each of these
clouds, the ratio of the stationary component of the 6.4 keV line
produced by the protons, $F^p_{6.4}$, to the time variable
component at its peak value from irradiation by primary X-rays,
$F^X_{6.4}$. To do this we use the observed ratio
$F^p_{6.4}/F^X_{6.4}=0.1$
 for the Bridge as it follows  from the
XMM-Newton data  \citep{ponti}. For the clouds G0.11$-$0.11 and
Sgr B2 this ratio as a function of the diffusion coefficient $D$
is shown in figure \ref{fratio}.

One can see that protons can contribute to the total 6.4 keV flux
from Sgr B2 if the diffusion is large enough, $D\gtrsim 10^{27}$
cm$^2$ s$^{-1}$. Then the expected   stationary flux should be one
order of magnitude less than the observed 6.4 keV emission from
Sgr B2  near its maximum in 2000. For small values of $D$ there is
no chance to observe 6.4 keV emission from Sgr B2 when the X-ray
front has crossed the cloud. For the cloud G\,0.11$-$0.11 which
according to \citet{ponti} is relatively close to Sgr A$^\ast$ the
situation is different. The intensity of stationary component is
quite high almost independently of $D$ and, in principle, may be
detected in several years.

 \citet{ponti} estimated the front width of primary X-rays from
 a non-detection of 6.4 keV emission from the two molecular (the 20
and 50 km s$^{-1}$ clouds) with a mass more than $10^4M_\odot$
(\cite{tsuboi}) which are within 15 pc of Sgr A$^\ast$.
  \citet{ponti} assumed  the X-ray front had passed already
these clouds which were very close to Sgr A$^\ast$  and,
therefore,  they  do not shine anymore. From figure \ref{fratio}
it follows that in this case a stationary 6.4 keV component should
be seen after the front passage. We notice that the distances to
the clouds was estimated from the assumption that the envelopes of
nearby SN remnants interact with these clouds \citet{coil}. If
this is true then it is very surprising that fluxes of continuum
and line emission
  are not observed from these clouds at all (as expected in the model of \cite{yus1}). As
follows from \citet{byk} when a shock front of SN interacts with a
molecular cloud,  energetic electrons  generated at the
 shock produces an intensive flux of hard X-rays from the
cloud. So, it is very strange that in such a situation X-ray
emission is not observed at all from these two clouds if the
interpretation of cloud - SN interaction is correct. If one accept
this interpretation then very special conditions for high energy
particle propagation should be assumed around the clouds. Besides,
as follows from \citet{sofue} and \citet{sawa} is not easy to
determine the distances between these clouds and Sgr A$^\ast$. In
principle, the XMM-Newton data do not exclude also any stationary
component of the 6.4 keV flux from these clouds below the derived
upper limit.

\section{ Predicted Variations of the Sgr B2 eW in Near Future}
\begin{figure*}[ht]
\begin{center}
\FigureFile(160mm,80mm){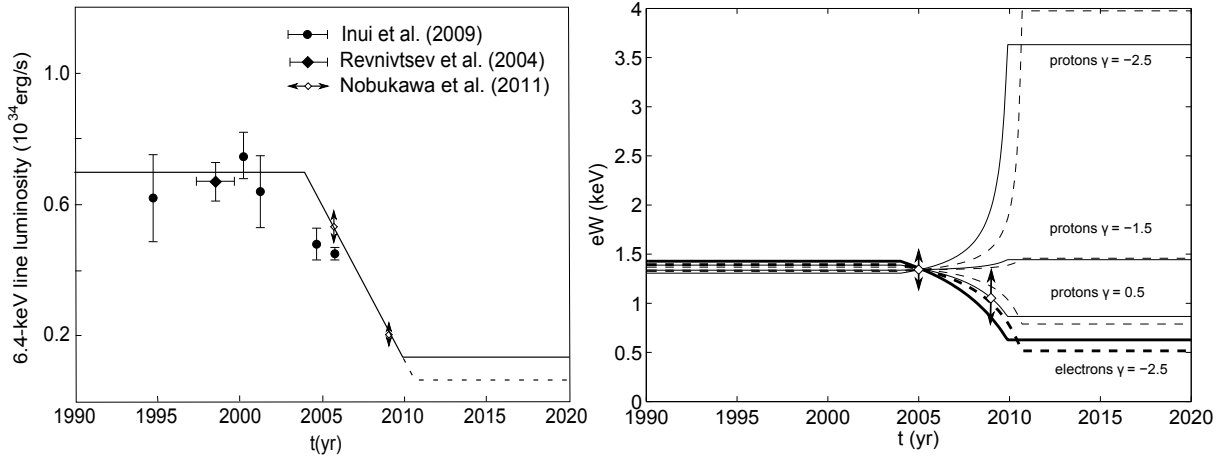}
\end{center}
\caption{{\it Left:} The evolution of the Fe K$\alpha$ line
luminosity and X-ray continuum as observed for Sgr B2. {\it
Right:} The possible evolution of Fe K$\alpha$ line equivalent
width. Dashed lines correspond to $F^p_{6.4}/F^X_{6.4} = 0.1$,
solid lines correspond to $F^p_{6.4}/F^X_{6.4} = 0.2$.
}\label{eqw_time}
\end{figure*}

Observations show that the flux of the 6.4 keV line emission from
Sgr B2 is rapidly decreasing with time (see the left panel of
figure \ref{eqw_time}). The question is whether we can find any
evidence for a possible stationary component of Sgr B2. In figure
\ref{eqw_time} (right panel) we presented the expected variations
of the Sgr B2 equivalent width when the flux generated by the
primary X-rays, $F^X_{6.4}$, is dropping down to the level 20\%
(solid lines) and 10\% (dashed lines) of the maximum value with
the rate shown in the left panel of the figure. The calculations
were done for protons with different spectral indexes $\gamma$ and
for electrons with $\gamma=-2.7$. One can see from the figure that
in the case if these particles are electrons the value of {\it eW}
decreases (almost independent of the electron spectral index, see
figure \ref{eqw_gamma}). In the case of protons the situation is
intricate: for soft proton spectra (negative $\gamma$) the value
of $eW$ should increase with time while for spectra with a
positive spectral index it drops down. However, production of
 spectra with a positive $\gamma$ in the Galaxy seems doubtful.
 In this figure we showed also the measured value of $eW$ for the
 years 2005 and 2009 (see \cite{nobukawa1}). Unfortunately, it is
 difficult to derive a time trend of the $eW$ variations because of
 relatively large error boxes.

These calculations show that the equivalent width should in
principle change if there is a component of Sgr B2 emission
generated by subrelativistic particles.  It follows from figure
\ref{eqw_time} that if the {\it eW} is decreasing with time than
the origin of impact component is due to electrons. In the
opposite case stationary component of 6.4 keV emission is produced
by subrelativistic protons. If future observations do not find any
time variation of the {\it eW} of the 6.4 keV line that will be a
strong evidence in favour of their pure photoionization origin.

 Recent Suzaku observations  may find the iron line emission
which is produced by subrelativistic particles \citep{fuku,
tsuru}. For the clumps G\,0.174$-$0.233 with $eW\simeq 950$ eV
they concluded that the  X-ray reflection nebula (XRN) scenario
was favored. On the other hand, for the clump 6.4 keV
G\,0.162$-$0.217 with $eW\simeq 200$ eV they assumed that the
emission from there was due to  low energy cosmic-ray electron
(LECRe). They found also that the $eW$ of the 6.4 keV emission
line detected in the X-ray faint region (non galactic molecular
cloud region)  is significantly lower than one expected in the XRN
scenario but higher than that of the LECRe model.  In this respect
we notice that for the spectrum of protons in the interstellar
medium of the GC with the spectral index $\gamma=0.5$, as derived
by \citet{dog_aa}, the $eW$ of emission produced by protons is
smaller than that of photoionization, that may explain these new
Suzaku results (see Figs. \ref{eqw_gamma} and \ref{eqw_time} for
the proton spectral index $\gamma=0.5$).

Future experiment can also distinguish the line origin from its
width. If electrons and X-rays generate a very narrow 6.4 keV line
with the width about 1 eV, the line produced by subrelativistic
protons is rather broad, $< 100$ eV (see \cite{dog4}). The
estimated width of the Fe K line for the model presented in
\citet{dog_pasj1} is about 40 eV.  If there is a noticeable proton
component of the 6.4 keV flux from the clouds, the width of the
line should broaden with time.

Measurements of the line with present X-ray telescopes contains
broadening which depends on photon statistics
 and calibration uncertainties. The energy resolution of CCD
detectors at 6 keV is $\sim$130 eV which can be decreased after
the de-convolution procedure (see \cite{koya2,ebi2}). However,
even with this procedure it is not easy to derive a true line
width from observations, if it is about 40 eV.  For more reliable
results a detector with a high energy resolution of ~eV such as
micro-calorimeter Astro-H/SXS is necessary.

\section{Conclusion}

We investigated parameters of the K$\alpha$ line emission from the
molecular clouds in the GC when it is excited by a flux of
subrelativistic protons. These protons are generated by accretion
onto the super-massive black hole. We concluded that:
\begin{itemize}
\item If these protons are generated by accretion processes they
produce a quasi-stationary component of 6.4 keV line and continuum
hard X-ray emission from molecular clouds in the GC because of
their very long lifetime. In this situation two components of
X-ray radiation should be observed: a time variable emission due
to photoionization by primary X-ray photons emitted by Sgr
A$^\ast$ and a quasi-stationary component generated by proton
impact.
\item Since the cross-sections of continuum and the iron line
production are different for these two processes, we expect that
they can be distinguished from the analysis of the equivalent
width of the iron line and we can predict time variations of {\it
eW}
 when the photoionization flux drops down after the passage of  X-ray
 front injected by Sgr A$^\ast$.
\item Whether or not the stationary component excited by protons
can be observed, depends on a distance of a cloud from Sgr
A$^\ast$ and the coefficient of spacial diffusion in the GC
medium. For the cloud G\,0.11$-$0.11 which is relatively close to Sgr
A$^\ast$ we expect to observe in a few years a stationary
component of the 6.4 keV emission  at the level $\lesssim$ 10\%
from its present value. For the cloud Sgr B2 the situation is more
intricate. If the diffusion coefficient $D\gtrsim 10^{27}$
cm$^2$s$^{-1}$ then the expected stationary flux should be about
10\% of its level in 2000. In the opposite case the line intensity
from Sgr B2 should drop down to zero because the protons do not
reach the cloud.
\item When the front of primary X-rays is passing through the
clouds, the density of primary X-ray photons decreases and the
relative contribution of the stationary iron line emission, if
presents, into the total flux increases. Therefore, parameters of
the emission from clouds changes with time. We expect that the
spectrum of charged particles generating the stationary component
can be derived from time variability of the line equivalent width.
\item We
showed that the equivalent width of the iron line excited by
charged particles depends of their charge composition and spectral
index $\gamma$. The equivalent width of K$\alpha$ line generated
by electrons depends weakly on $\gamma$, and it varies from $\sim
250$ eV for soft spectra to $\sim 500$ eV for hard electron
spectra. In the case of protons the width variations are
significant reaching its maximum for very soft proton spectra.
\item If future observations  find any
time variation of the {\it eW} of the 6.4 keV line, then in the
case of decrease the impact line component is produced by
electron, in the opposite case - by subrelativistic protons.
\end{itemize}

\vspace{5 mm}The authors are grateful to Vincent Tatischeff for
the data shown in figure~1 and to the unknown referee  who made
much to improve the text.

 VAD and DOC are partly supported by  the NSC-RFBR Joint Research Project RP09N04 and
09-02-92000-HHC-a. This work is also supported by Grant-in-Aids
from the Ministry of Education, Culture, Sports, Science and
Technology (MEXT) of Japan, Scientific Research A, No. 18204015
(KK). MN is supported by JSPS Research Fellowship for Young
Scientists. KSC is supported by a GRF grant of Hong Kong
Government 7011/10p.

\vspace{8 mm}


\end{document}